\documentclass{article}
\usepackage[a4paper, left=1.5cm, right=1.5cm, top=2.5cm, bottom=2.5cm]{geometry}
\usepackage{graphicx} 
\usepackage{amsmath}
\usepackage{amssymb}
\usepackage{xcolor}
\usepackage{authblk} 
\usepackage{cite}

\title{Evolving the Loeb Scale}

\author[1]{Oem Trivedi\thanks{oem.trivedi@vanderbilt.edu}}
\author[2]{Abraham Loeb\thanks{aloeb@cfa.harvard.edu}}

\affil[1]{Department of Physics and Astronomy, Vanderbilt University, Nashville, TN 37235, USA}
\affil[2]{Astronomy Department, Harvard University, 60 Garden St., Cambridge, MA 02138, USA}

\date{\today}

\begin{document}

\maketitle

\begin{abstract}
We develop a differential formulation of the Loeb Scale that extends the original static framework into a radially evolving, real–time classification scheme for interstellar objects. By promoting each anomaly metric to a function of heliocentric distance and introducing a relaxation equation for the effective score, our method incorporates memory, hysteresis and predictive capability. This allows us to have early, stable forecasts of an object's eventual Loeb level based on sparse data obtained at large distances, which is more helpful to quantify its true nature when near Earth.
\end{abstract}

\section{Introduction}

Over the past decade, the discovery of interstellar objects (ISOs) has transformed our understanding of the diversity of bodies that traverse the Solar System. The identification of 1I/`Oumuamua in 2017 \cite{oumumeech2017brief}, 2I/Borisov in 2019 \cite{borguzik2020initial} and most recently 3I/ATLAS in 2025 \cite{seligman2025discovery} has opened an entirely new window into the study of extrasolar planetesimals. While 2I/Borisov behaved as a conventional comet, the anomalous characteristics of 1I/`Oumuamua and 3I/ATLAS \cite{oum1loeb2022possibility,oum2forbes2019turning,oum3siraj2022mass,oum4siraj20192019,oum5bialy2018could,3i1loeb20253i,3i2a2hibberd2025interstellar,3i3tde2025assessing,3i4a3loeb2025intercepting,3i5hopkins2025different} show the possibility that the Solar System may occasionally be visited by objects that deviate significantly from the physical and dynamical properties of familiar comets and asteroids. With the imminent operations of the Vera C. Rubin Observatory \cite{vc1thomas2020vera,vc2blum2022snowmass2021,vc3sebag2020vera}, detection rates of ISOs are expected to rise by one to two orders of magnitude, making it imperative to develop quantitative tools that can rapidly assess the nature of newly discovered objects and discriminate between ordinary interstellar debris and bodies exhibiting potentially technological signatures.
\\
\\
As the catalog of ISOs grows, so too does the need to evaluate not only their scientific significance but also the extent to which they may pose a hazard to Earth. Motivated by this challenge, the Loeb Scale was introduced as a structured ten-level classification scheme that ranks objects according to the degree of anomaly they exhibit relative to natural icy rocks \cite{eldadi2025loeb}. Much like the Kardashev scale provides us with a classification for the energy capacities of civilizations \cite{kar1kardashev1964transmission,kar2kardashev1980strategies,kar3cirkovic2016kardashev,kar4gray2020extended,kar5namboodiripad2021predicting,kar6sonia2022civilizations}, the Loeb Scale offers a unified language for characterizing potential interstellar artifacts. It considers a wide variety of objects, ranging from objects entirely consistent with natural origins (Level 0) to those whose behavior may indicate artificial construction or even constitute a technological threat (Levels 8–10). While this framework has provided an essential conceptual foundation, the increasing pace of ISO discoveries demands methods that can evaluate their Loeb classification continuously as new observations accumulate during their passage through the Solar System.
\\
\\
A full mathematical formulation of the Loeb Scale was established in Ref. \cite{loe1trivedi2025quantitative}, providing a quantitative mapping from observed anomalies to a continuous score and subsequent discrete level assignment. One key observation though is that this formulation remains fundamentally static, depending on measurements obtained at a single epoch and offering no means to incorporate the evolving physical and dynamical characteristics of an ISO as it approaches the inner Solar System. Because most ISOs are detected at large heliocentric distances where observational uncertainties are substantial, a static evaluation may poorly reflect the eventual classification once richer datasets become available near Earth. This motivates the development of a differential, radially evolving version of the Loeb Scale that updates continuously with incoming data naturally incorporates memory of sustained anomalies and forecasts the likely Loeb level by the time the object reaches Earth's vicinity. This is what we aim to achieve in this work and we organize it as follows. In section 2 we review the mathematical structure of the Loeb Scale and in section 3 we introduce the differential evolution equation governing its radial behavior. In section 4 we discuss caveats and limitations of this formulation and in section 5 we summarize our conclusions.
\\
\\
\section{Mathematical Foundations of the Loeb Scale}
In order to formulate a differential generalization of the Loeb scale, it is useful to summarize the mathematical framework of the scale itself as developed in our previous work \cite{loe1trivedi2025quantitative}. The Loeb scale is a ten-level classification scheme for the technosignature significance of interstellar objects, ranging from fully natural bodies at level 0 to confirmed existential threats at level 10. The purpose of the scale is to provide a reproducible and quantitative mapping from observational anomalies to a well defined integer level that reflects both the physical character of the object and its potential technological implications. To achieve this, one begins by defining a set of normalized anomaly metrics that encode the degree to which a given observable departs from expectations for natural Solar System populations.
\\
\\
Note that here each metric is constructed from raw measurements and is transformed into a normalized variable $m_i \in [0,1]$, where $m_i=0$ indicates full consistency with natural behavior and $m_i=1$ represents a maximally anomalous, technologically suggestive or extreme value. The metrics include non-gravitational acceleration anomaly $A$, spectral or compositional anomaly $B$, shape or lightcurve anomaly $C$, albedo or surface-weathering anomaly $D$, trajectory or targeting improbability $E$, electromagnetic signal significance $F$, operational or behavioral indicators $G$ and optionally an impact-risk factor $H$ for differentiating between upper levels. Each metric is computed from a raw observable and mapped into the normalized range via monotonic transforms and calibrated clamping functions. To begin, we would briefly summarize the various metrics considered, starting with the non-gravitational acceleration anomaly which begins with the raw value
\begin{equation}
A_{\mathrm{raw}} = \log_{10}\left(\frac{a_{\mathrm{obs}}}{a_{\mathrm{ref}}}\right)
\end{equation}
where $a_{\mathrm{obs}}$ denotes the measured non-gravitational acceleration and $a_{\mathrm{ref}}$ is a reference value chosen to represent nominal cometary behavior and the raw quantity is mapped into $[0,1]$ through
\begin{equation}
A = \mathrm{clamp}\left(\frac{A_{\mathrm{raw}} + 2}{4},\,0,\,1\right)
\end{equation}
where the constants shift and scale the logarithmic range so that typical cometary accelerations yield values near $A\approx 0.5$.
\\
\\
The spectral anomaly metric $B$ compares observed spectra, gas production rates and line ratios to empirical population distributions of cometary species. If $\chi^{2}_{\mathrm{mismatch}}$ denotes a measure of deviation between the observed spectrum and the best-fit natural template then one may define the continuous mapping
\begin{equation}
B = \mathrm{clamp}\left(\frac{\chi^{2}_{\mathrm{mismatch}}}{\chi^{2}_{\mathrm{mismatch}} + K_B},\,0,\,1\right)
\end{equation}
where $K_B$ is a tunable constant controlling the sensitivity and a more refined construction uses the population percentile of each measured quantity, which can be defined as
\begin{equation}
s_x = 1 - 2 \min\left(F_{\mathrm{pop}}(x_{\star}),\,1 - F_{\mathrm{pop}}(x_{\star})\right)
\end{equation}
where we note that $F_{\mathrm{pop}}$ is the cumulative distribution function for the relevant cometary dataset. For censored measurements with upper limits, one replaces $F_{\mathrm{pop}}$ by the corresponding survival function. Multiple indicators are used here which are combined as a weighted sum of their rarity scores and mapped to $[0,1]$ by a rational transform of the form
\begin{equation}
B = \mathrm{clamp}\left(\frac{\sum_k \alpha_k s_{x_k}}{\sum_k \alpha_k s_{x_k} + K_B},\,0,\,1\right)
\end{equation}
The shape anomaly $C$ is derived from the inferred aspect ratio $R$ of the body, which ends up using
\begin{equation}
C = \mathrm{clamp}\left(\frac{\log_{10}(R)}{\log_{10}(R_{\max})},\,0,\,1\right)
\end{equation}
where $R_{\max}$ is a maximum reference ratio chosen to encapsulate the upper tail of plausible natural shapes. The albedo anomaly $D$ is constructed relative to the two-Rayleigh mixture distribution that describes the empirical albedo distribution of small Solar System bodies. If the mixture probability density is denoted by
\begin{equation}
p_{2R}(p_V) = f_D\,\mathrm{Ray}(p_V; d) + (1 - f_D)\,\mathrm{Ray}(p_V; b)
\end{equation}
with the Rayleigh components
\begin{equation}
\mathrm{Ray}(x;\sigma) = \frac{x}{\sigma^{2}} \exp\left(-\frac{x^{2}}{2\sigma^{2}}\right)
\end{equation}
and empirically fitted parameters $f_D$, $d$, and $b$, then the rarity of an observed albedo $p^{\star}_V$ is quantified through the two-sided tail probability
\begin{equation}
p_{\mathrm{tail}}(p^{\star}_V) = \min\left(\int_0^{p^{\star}_V}p_{2R}(p_V)\,dp_V,\,\int_{p^{\star}_V}^{\infty}p_{2R}(p_V)\,dp_V\right)
\end{equation}
and the normalized albedo anomaly is
\begin{equation}
s_{\mathrm{albedo}} = 1 - 2 p_{\mathrm{tail}}(p^{\star}_V)
\end{equation}
This is then converted to $D$ by
\begin{equation}
D = \mathrm{clamp}\left(\frac{s_{\mathrm{albedo}}}{s_{\mathrm{albedo}} + K_D},\,0,\,1\right)
\end{equation}
The trajectory anomaly $E$ is based on the improbability of the arrival geometry under an isotropic flux of incoming interstellar objects and if $p$ denotes this probability, one defines
\begin{equation}
E = \mathrm{clamp}\left(\frac{-\log_{10}(p)}{X},\,0,\,1\right)
\end{equation}
where $X$ is a scaling parameter that tunes the sensitivity of the metric to rare arrival trajectories. One can also define the electromagnetic signal score $F$ and operational behavior metric $G$, which are constructed from monotonic transforms of narrowband signal-to-noise ratios, modulation properties, maneuvering residuals or sub-object detections. Here each are mapped into the unit interval using logistic or rational functions and the impact-risk factor $H$ is defined in terms of impact probability and kinetic energy, normalized so that objects posing negligible risk satisfy $H \approx 0$ and impactors with catastrophic energy yield $H \approx 1$.
\\
\\
Having defined all metrics $m_i$, the Loeb scale consolidates them into a single composite anomaly score $S \in [0,1]$ by combining linear contributions and pairwise synergies. Here we have the linear contribution being given as
\begin{equation}
S_{\mathrm{lin}} = \sum_{i} w_i m_i
\end{equation}
where the weights $w_i$ satisfy $w_i \ge 0$ and $\sum_i w_i = 1$ and to capture the fact that distinct anomalies reinforce each other, one includes interaction terms
\begin{equation}
S = \sum_{i} w_i m_i + \sum_{i<j} w_{ij} m_i m_j
\end{equation}
where $w_{ij}$ are small, tunable coefficients restricted to physically motivated pairs. Note that this composite score increases more strongly when multiple independent anomalies co-occur, reflecting heightened suspicion.
\\
\\
The composite score is mapped to the integer Loeb levels via calibrated thresholds where, for example, one may assign level 0 for $S<0.20$, level 1 for $0.20 \le S < 0.35$, level 2 for $0.35 \le S < 0.50$ and so forth, with the critical threshold for formal technosignature consideration placed at $S \approx 0.60$ corresponding to level 4. At the upper end scores $S\ge 0.995$ correspond to level 10, which indicates a confirmed artificial object on an Earth-impact trajectory with globally catastrophic consequences. Uncertainty propagation follows directly from the measurement uncertainties of each metric. In a first-order approximation, if $\sigma_{m_i}$ denotes the error of metric $m_i$, then the variance of $S$ is
\begin{equation}
\sigma_S^2 \approx \sum_{i}\left(w_i + \sum_{j\neq i} w_{ij} m_j\right)^2 \sigma_{m_i}^2
\end{equation}
although a full Monte Carlo propagation of the metric distributions is recommended for robust communication. By treating the metrics as functions of heliocentric distance and introducing a dynamical evolution equation for the effective score, one can extend the static mapping into a continuous, real-time classification scheme that evolves with observational data. The remainder of this work develops such a differential extension by promoting the composite score $S$ to a radially evolving quantity and then coupling it to parametric models for the radial dependence of each anomaly metric.
\\
\\
\section{Evolving the Loeb Scale}
In order to develop a dynamical formulation of the Loeb scale that updates continuously as observational data evolve, it is natural to promote the composite anomaly score into a radially dependent quantity. The instantaneous Loeb score already possesses an explicit mathematical definition in terms of the anomaly metrics evaluated at a single epoch and if $m_i(r)$ denotes the normalized value of metric $i$ at heliocentric distance $r$, then the instantaneous score is given by
\begin{equation*}
S_{\mathrm{inst}}(r) = \sum_{i} w_i m_i(r) + \sum_{i<j} w_{ij} m_i(r)m_j(r)
\end{equation*}
where the constants $w_i$ and $w_{ij}$ denote the linear and pairwise interaction weights introduced earlier. This expression reduces to the static Loeb score when evaluated at a single fixed value of $r$, but for an object whose properties evolve through its solar encounter, one generally obtains different values of $S_{\mathrm{inst}}$ as more data accumulate at different distances and the quantity $S_{\mathrm{inst}}(r)$ therefore represents the raw anomaly score inferred directly from the most recent measurements, without any smoothing, averaging or persistence of past information.
\\
\\
A fundamental limitation of using $S_{\mathrm{inst}}(r)$ directly is that real observational data are uneven in quality and cadence  and brief anomalous measurements at isolated radii may spuriously elevate or depress the Loeb classification. To provide a mathematically stable and physically interpretable alternative, one may define an effective anomaly score $S_{\mathrm{eff}}(r)$ that evolves gradually toward the instantaneous value without matching it immediately. This prescription introduces a form of dynamical memory into the classification and reflects the expectation that sustained anomalies are more significant than transient ones and a simple formulation achieving this goal is a radial first order relaxation equation,
\begin{equation} \label{diffeq}
\frac{d S_{\mathrm{eff}}}{dr} = \frac{S_{\mathrm{inst}}(r) - S_{\mathrm{eff}}(r)}{L}
\end{equation}
where $L$ is a characteristic relaxation length scale measured in astronomical units. The parameter $L$ determines the responsiveness of the dynamical score and note that if $L$ is small then $S_{\mathrm{eff}}(r)$ closely traces $S_{\mathrm{inst}}(r)$ and the classification reacts rapidly to new information. For larger $L$, the evolution becomes more inertial and significant changes in heliocentric distance are required before the effective score moves appreciably toward the instantaneous value. This structure induces a natural hysteresis as short lived fluctuations in individual anomaly metrics do not immediately alter the classification and only sustained departures from natural expectation generate long term changes in the effective level.
\\
\\
To evaluate $S_{\mathrm{eff}}(r)$ one must first specify the functional dependence of each anomaly metric $m_i(r)$ on heliocentric distance and because new data typically arrive sparsely and with significant uncertainty, it is useful to describe these metrics using simple parametric forms that can be continuously updated as improved constraints become available. A representative example is provided by the non-gravitational acceleration anomaly. If $a_{\mathrm{obs}}(r)$ is the measured nongravitational acceleration and $a_{\mathrm{nat}}(r)$ a reference value predicted by a natural sublimation model, then one may write
\begin{equation}
A(r) = \mathrm{clamp}\left(\frac{\log_{10}[a_{\mathrm{obs}}(r)/a_{\mathrm{nat}}(r)] + C_A}{D_A}, 0, 1\right)
\end{equation}
with the reference model parametrized as
\begin{equation}
a_{\mathrm{nat}}(r) = a_0\left(\frac{1}{r}\right)^n \Theta(r - r_{\mathrm{ice}})
\end{equation}
note here that $n$ controls the steepness of the sublimation response, $r_{\mathrm{ice}}$ defines the characteristic activation radius of relevant volatiles, and $\Theta$ is a smoothed step function that switches on activity as $r$ decreases. This formulation keeps the anomaly metric well defined even when only a few acceleration measurements exist and the parameters $(a_0,n,r_{\mathrm{ice}})$ may be refined as the object is observed over a wider range of distances.
\\
\\
The spectral anomaly metric may be expressed as a sigmoid function in $r$, which goes towards reflecting the onset of gas emission or unusual chemical signatures as the object receives greater insolation. A convenient representation in this case would be
\begin{equation}
B(r) = B_{\max}\left[1 + e^{(r - r_{\mathrm{crit}})/\Delta r}\right]^{-1}
\end{equation}
where $r_{\mathrm{crit}}$ denotes a characteristic activation radius and $\Delta r$ measures the sharpness of the transition. This captures early or delayed onset behavior and naturally accommodates nondetections at large heliocentric distances. There are also some metrics which vary only weakly with $r$ but have uncertainties that shrink as additional measurements become available. The shape anomaly $C(r)$ and albedo anomaly $D(r)$ fall into this category and a simple model for the radial dependence of their uncertainties is
\begin{equation}
\sigma_C(r) = \sigma_{C,0}\exp[-N_{\mathrm{LC}}(r)/N_0]
\end{equation}
where $N_{\mathrm{LC}}(r)$ is the cumulative number of lightcurve measurements obtained up to distance $r$ and $N_0$ is a scale factor controlling the reduction rate. The mean values of $C(r)$ and $D(r)$ may be treated as approximately constant, but the shrinking uncertainty allows the instantaneous score $S_{\mathrm{inst}}(r)$ to become more accurate as the object approaches the inner Solar System.
\\
\\
The trajectory anomaly metric is particularly sensitive to improved orbit determination as one would expect and so if $p(r)$ is the isotropic arrival probability based on the best fit orbit solution at distance $r$, one may express it as
\begin{equation}
p(r) = p_{\mathrm{iso}}\exp[-\kappa Q(r)]
\end{equation}
where $Q(r)$ quantifies how geometrically unusual the fitted orbit is relative to the isotropic assumption and the anomaly metric is then
\begin{equation}
E(r) = \mathrm{clamp}\left(-\frac{\log_{10}p(r)}{X},0,1\right)
\end{equation}
with $X$ the scaling parameter introduced earlier. As astrometric uncertainties shrink with additional observations, the value of $Q(r)$ may rise rapidly which ends up producing a corresponding radial growth in $E(r)$ if the object's trajectory is unexpectedly close to a significant Solar System target. Once the radial dependence of all metrics has been specified, the instantaneous score can then follow from
\begin{equation}
S_{\mathrm{inst}}(r) = \sum_i w_i m_i(r) + \sum_{i<j} w_{ij}m_i(r)m_j(r)
\end{equation}
and the effective score is obtained by integrating \eqref{diffeq} and the resulting function $S_{\mathrm{eff}}(r)$ may then be compared against the level thresholds, with hysteresis introduced by requiring that the effective score remain above (or below) a threshold over a finite radial interval before a change in classification is adopted. This ensures that transitions between levels arise from persistent radial trends rather than isolated data points or transient anomalies. The differential formulation here incorporates the Loeb score as an evolving quality and quantity of observational data in a natural way and takes a principled notion of memory, enabling the classification to reflect sustained evidence of anomalous behavior. This makes the framework well suited for real time tracking of newly discovered interstellar objects and suggests practical applications such as automated monitoring pipelines that update the object's effective Loeb score as new observations become available.

\section{Operational directions and Caveats}
The differential formulation of the Loeb scale that we developed above admits a concrete operational implementation once an interstellar object is detected at large heliocentric distance, for example near the Kuiper Belt. At the detection radius $r_{\mathrm{det}}$, the anomaly metrics $m_i(r)$ are only weakly constrained, but one can already construct parametric models $m_i(r;\theta_i)$ that could take in both the limited data and prior expectations for natural objects. In this representation, the instantaneous anomaly score becomes a function of the heliocentric distance and the parameter set
\begin{equation}
S_{\mathrm{inst}}(r;\theta) = \sum_i w_i m_i(r;\theta_i) + \sum_{i<j} w_{ij}\,m_i(r;\theta_i)\,m_j(r;\theta_j)
\end{equation}
where $\theta \equiv \{\theta_i\}$ denotes the full set of metric parameters. The effective Loeb score $S_{\mathrm{eff}}(r)$ is then governed by the evolution equation described in \eqref{diffeq} and subject to an initial condition $S_{\mathrm{eff}}(r_{\mathrm{det}}) = S_{\mathrm{det}}$, where $S_{\mathrm{det}}$ is obtained from the initial data. For a given choice of parameters $\theta$, one can write the formal solution of the first order evolution as
\begin{equation}
S_{\mathrm{eff}}(r;\theta) = \mathcal{K}(r,r_{\mathrm{det}})\,S_{\mathrm{det}} + \int_{r_{\mathrm{det}}}^{r} \mathcal{K}(r,r')\,\mathcal{F}(r';\theta)\,dr'
\end{equation}
where $\mathcal{K}(r,r')$ is the Green’s function associated with eq. Loeb and $\mathcal{F}(r;\theta)$ is a source term proportional to $S_{\mathrm{inst}}(r;\theta)$. The explicit forms of $\mathcal{K}$ and $\mathcal{F}$ follow straightforwardly from \eqref{diffeq}. This expression shows that the effective score at any future radius $r$ is a weighted combination of the initial score and a radial integral of the instantaneous anomaly, with recent values of $S_{\mathrm{inst}}$ contributing more strongly than distant ones along the trajectory.
\\
\\
To forecast the Loeb scale near Earth, one can evaluate the distribution of $S_{\mathrm{eff}}(r_{\oplus};\theta)$ at the heliocentric distance $r_{\oplus} \approx 1\,\mathrm{au}$ and one can note here that at early times, when the observational constraints are weak, the parameters $\theta$ are described by broad priors or posteriors $P(\theta|{\cal D}_{\mathrm{det}})$ conditioned on the initial dataset ${\cal D}_{\mathrm{det}}$ at $r_{\mathrm{det}}$. One may then define the predictive distribution
\begin{equation}
P\big(S_{\mathrm{eff}}(r_{\oplus})\big|{\cal D}_{\mathrm{det}}\big) = \int d\theta\,P(\theta|{\cal D}_{\mathrm{det}})\,\delta\big(S_{\mathrm{eff}}(r_{\oplus};\theta) - s\big)
\end{equation}
which can be estimated in practice by Monte Carlo sampling of $\theta$, propagating each realization forward in radius and recording the resulting $S_{\mathrm{eff}}(r_{\oplus};\theta)$. The mean and credible intervals of this distribution provide an operational forecast for the effective Loeb score at Earth long before the object reaches the inner Solar System. As additional data ${\cal D}_{r}$ are collected at intermediate radii $r$, the parameter distribution is updated to $P(\theta|{\cal D}_{\mathrm{det}},{\cal D}_{r})$ and the forecast for $S_{\mathrm{eff}}(r_{\oplus})$ is recomputed. The continuous dependence on $r$ ensures that these updates can be performed at any stage of the approach without discontinuities.
\\
\\
In a simplified deterministic implementation, one may use a best fit parameter set $\hat{\theta}$ obtained from the current data and compute the corresponding trajectory $S_{\mathrm{eff}}(r;\hat{\theta})$ as a function of radius and the predicted Loeb score at Earth is then simply
\begin{equation}
S_{\oplus} \equiv S_{\mathrm{eff}}(r_{\oplus};\hat{\theta})
\end{equation}
while the uncertainty can be approximated by linear error propagation or, more robustly, by sampling around $\hat{\theta}$ within its covariance matrix. This procedure gives us an evolving forecast $S_{\oplus}$ that is updated each time new measurements refine the metrics, for example when improved astrometry tightens $E(r)$ or when new spectra constrain $B(r)$. Because the effective score depends on a radial integral of $S_{\mathrm{inst}}(r)$, transient spikes in individual metrics at isolated radii contribute only modest corrections, preserving the stability of the forecast unless persistent anomalies develop.
\\
\\
Operationally, an automated “Loeb monitor” for an ISO would therefore consist of a sequence of steps which can be summarized as follows. At each new observation epoch, the metrics $m_i(r)$ are updated and their parametric forms $m_i(r;\theta_i)$ refitted,  the instantaneous score $S_{\mathrm{inst}}(r;\theta)$ is recomputed along the future trajectory, the equation \eqref{diffeq} is integrated from the current radius to $r_{\oplus}$ to obtain a new prediction for $S_{\mathrm{eff}}(r_{\oplus};\theta)$ and the resulting distribution or median value is mapped to an anticipated Loeb level via the same thresholding scheme used for static objects. This process can be repeated as frequently as new data become available, ensuring that the classification forecast incorporates the latest measurements while preserving the hysteresis and smoothing inherent in the differential formulation.
\\
\\
An important feature of this framework is its adaptability as the global population of interstellar objects becomes better characterized. As more ISOs are discovered and analyzed, the empirical distributions underlying the anomaly metrics will narrow and the priors on $\theta$ will become more informative as  example, the reference distribution for non-gravitational accelerations, spectral ratios and albedo values will be derived from a larger and more diverse sample of natural interstellar bodies. In this regime, the mapping from observables to rarity scores and hence to anomaly metrics will sharpen and the same measured properties may lead to higher or lower anomaly scores than in the early days of ISO science. This implies that for any given object, its inferred anomaly profile $m_i(r;\theta_i)$ and consequently its effective score $S_{\mathrm{eff}}(r)$, may acquire a mild time dependence in retrospect as the community’s understanding of “normal” interstellar behavior improves. An object that initially appeared anomalous under broad, poorly constrained distributions might later be recognized as typical or conversely, it may move further into the tails of a better measured population, revealing its anomalous nature more starkly.
\\
\\
Another conceptual caveat arises when applying the Loeb scale or its differential generalization that we have developed here, to objects detected near Earth for which no interstellar origin has been established. The Loeb scale is designed to quantify anomalies relative to natural interstellar bodies and thus uses ISO-based distributions as its baseline and for objects in Earth orbit or cis-lunar space that might represent non-human technologies, the appropriate comparison class is not natural comets or asteroids but rather human-made spacecraft and debris. In such cases, the anomaly metrics must be redefined so that $m_i=0$ corresponds to full consistency with known human technologies and $m_i=1$ corresponds to behavior incompatible with any catalogued class of artificial objects. The mathematical structure of the composite score and the differential evolution described in \eqref{diffeq} can be retained, but the underlying probability distributions and normalization conventions must be replaced by those derived from the ensemble of human-made devices. This adaptation ensures that the interpretation of “anomaly” remains meaningful as what is highly unusual for a natural ISO may be entirely mundane for a human satellite and vice versa.
\\
\\

\section{Conclusions}

The differential formulation of the Loeb scale presented in this work extends the original static framework into a continuously evolving system capable of assimilating observational data across a full heliocentric trajectory. By promoting each anomaly metric to a radial function and introducing an effective score that responds gradually to new information, the Loeb scale can now produce stable and predictive classifications long before an interstellar object reaches the inner Solar System. The formal structure developed here establishes a clear mathematical relationship between metric evolution, instantaneous anomaly significance, and the smoothed effective score, thereby providing a natural and physically motivated means of incorporating memory, hysteresis and forecast capability into the classification scheme.
\\
\\
This reformulation also enables practical forward modeling once an object is detected at large heliocentric distance. Through parametric representations of the metrics and predictive evaluations of the instantaneous anomaly score, one can integrate the Loeb evolution equation inward and estimate the future classification at Earth’s distance, complete with uncertainty quantification arising from parameter posteriors and the operational pipeline developed in this framework is flexible enough to accommodate hardly available early data, increasingly precise measurements in the inner Solar System and retrospective updates as the empirical understanding of natural interstellar populations improves. 
\\
\\
Looking ahead, the dynamical formulation opens several possibilities as given the notion that large surveys discover growing numbers of interstellar objects, the anomaly metrics themselves will sharpen which will end up enabling the Loeb scale to become increasingly predictive and discriminating. One may envisage automated monitoring systems that continuously ingest astrometric, photometric, spectroscopic, and radar data to produce evolving anomaly forecasts for every newly detected ISO. More ambitiously, the same mathematical structure could be adapted into an early warning system for potential technological objects, guiding observational priorities and even future intercept missions. As the Solar System becomes a richer laboratory of interstellar visitors and as the frontier of anomalous object science expands, the differential Loeb scale may serve as a unifying, quantitative language for identifying natural outliers, evaluating technosignatures and ultimately perhaps also informing us about the search for extraterrestrial technologies across multiple observational domains.
\\
\\
\section*{Acknowledgments}
OT was supported in part by the Vanderbilt Discovery Alliance Fellowship. AL was supported in part by the Galileo Project and  the Black Hole Initiative.

\bibliography{references}
\bibliographystyle{unsrt}

\end{document}